# Understanding Communication Patterns in MOOCs: Combining Data Mining and qualitative methods

**Rebecca Eynon[1,2], Isis Hjorth[1], Taha Yasseri[1], and Nabeel Gillani[1]**

[1]Oxford Internet Institute, University of Oxford, Oxford, UK

[2]Department of Education, University of Oxford, Oxford, UK

**Correspondence:**
Rebecca Eynon, rebecca.eynon@oii.ox.ac.uk
Oxford Internet Institute, University of Oxford, 1 St. Giles, Oxford OX1 3JS, UK.

**Abstract**

Massive Open Online Courses (MOOCs) offer unprecedented opportunities to learn at scale. Within a few years, the phenomenon of crowd-based learning has gained enormous popularity with millions of learners across the globe participating in courses ranging from Popular Music to Astrophysics. They have captured the imaginations of many, attracting significant media attention – with The New York Times naming 2012 "The Year of the MOOC." For those engaged in learning analytics and educational data mining, MOOCs have provided an exciting opportunity to develop innovative methodologies that harness big data in education.

**Introduction**

At these early stages of exploring how learning unfolds in large-scale learning environments, it is becoming clear that significant methodological challenges remain. In particular, we argue that qualitative or quantitative approaches are not, on their own, sufficient to extract meaningful insights into how people learn in these settings. We suggest that particularly constructive ways of addressing these challenges include the adoption of pragmatic research paradigms (Tashakkori and Teddlie, 1998), embracing multi-level exploration of data (Welser et al., 2008), informed by critical engagement with contemporary learning theory. We expand this argument further below, before offering a reflexive account of methodological approaches and analytical strategies that can be combined in mixed-method research designs to provide more rigorous conceptual understandings of learning at scale. The reflective discussion draws upon our own research into learning in MOOCs (Eynon et al., 2014; Gillani et al. 2014a, 2014b; Gillani and Eynon, 2014).

The significant interest in researching MOOCs from a range of different disciplines is partially due to the availability and abundance of digital trace data that are produced by learners in mass-scale online environments. MOOCs offer researchers fine-grained data collected on learners' participation and mutual interactions (e.g. Breslow et al., 2013; Kizilcec et al., 2013) that have never been available at such scales before. These "digital traces" were once virtually impossible to capture in campus-based contexts – such as the number of students that opened a textbook before the final exam, or which students spoke to which other students about the final problem set. The kinds of data available in digital learning environments have fuelled a proliferation of interdisciplinary research, bringing together computational and social scientists to collectively ask and answer questions about how learning happens in massive-scale online courses.

These interdisciplinary collaborations have revealed, however, that learning is indeed a complex process, and understanding it requires tools beyond advanced computational techniques. This is because learning cannot be gauged alone through behaviours codified by digital trace data; it is constituted by cognitive and sociological elements too. Illeris (2003) provides a very useful definition that captures the complexity of the learning process. He suggests the process of learning can be viewed as:

> An entity which unites a cognitive, an emotional and a social dimension into one whole. It combines a direct or mediated interaction between the individual and its material and social environment with an internal psychological process of acquisition. Thus, learning always includes an individual and a social element, the latter always reflecting current societal conditions, so that the learning result has the character of an individual phenomenon which is always socially and societally marked. (p. 227)

When an individual learns something, it is both their behaviour and their experience of that behaviour that is important; and this experience is shaped by the context of which they are part of, which can involve other people. Considering the learning environments and affordances of MOOCs, it is clear that the 'social' and communication form important aspects of such contexts. The reason we emphasise the role of communication in MOOCs, and what this means for learning, is because when one considers what MOOCs can potentially offer learners that previous incarnations of open education initiatives have not, we argue that MOOCs are unique in the way that they offer an opportunity for thousands of learners from diverse geographical locations with varied experience to participate and collaborate with each other without physical presence.

In saying this, we are not claiming that all learning is social, as learning can occur in a variety of ways, both through activities that support the acquisition of information and knowledge as well as more collaborative and participative approaches (Sfard, 1998). However, the social element of Massive Open Online Courses is an important aspect to consider in learning. Indeed, from the significant amount of research in online learning, there are already a range of social constructivist and social-cultural perspectives that can be utilised (e.g. Goodman and Dabbish, 2011; Lave and Wenger, 1991; Siemens 2005; Stahl et al., 2006). Furthermore, recent studies have suggested that some students in MOOCs spend as much or more time using the discussion forums as they do viewing lectures or doing homework (Seaton et al., 2013), highlighting the need to also explore these kinds of activities.

Taking the above into account, we argue that in order to capture learning in mass-scale crowd-based environments, a mixed method approach is required which combines data mining with a wide set of social science techniques that are primarily qualitative in nature. These include methods such as observation, interviews and surveys, more traditionally used in education research. In promoting such an approach to research, we equally advocate for the explicit building upon existing work in the fields of education, learning and technology, resisting ahistorical approaches to studying this area.

**Methodological approaches to understanding communication patterns in MOOCs**

Mixed method research is challenging because of the pronounced differences in the epistemological underpinnings of research employing data mining and research associated with qualitative observation and interviews. We believe a constructive way of addressing this issue is to adopt a pragmatic paradigm, where the primary attention is given to the research question asked, as opposed to holding a particular allegiance to a philosophy or methodology when carrying out the research (Brannen, 1992; Hammersley, 1992). A key consequence of ascribing to the pragmatic paradigm is that the focus becomes identifying and critically engaging with the most suitable analytical and methodological techniques to answer specific research questions; regardless of whether they are traditionally viewed as quantitative or qualitative in nature.

The pragmatic paradigm has a number of key characteristics including the use of: 1) both qualitative and quantitative methods; 2) deductive and inductive logic; 3) objective and subjective viewpoints; 4) the important role of values when interpreting results; 5) the acceptance of choosing explanations of the research that produce desired outcomes and; 6) the exploration of causal linkages, but under the acknowledgement that while an attempt will be made to make linkages, they may not be defined precisely as data can lead to a number of explanations. Thus, these kinds of explanations will reflect researchers' values systems (Tashakkori and Teddlie, 1998:23).

In line with a pragmatic paradigm in MOOC research, multiple methods are not employed with the objective to reach neat triangulation of findings, nor as a means to use one method simply to validate the other findings. Instead, all methods are given equal value to ultimately illuminate how people learn and interact in MOOCs. The metaphor of crystallisation in bringing data sources together where each method produces information that provides one aspect on the problem (Ellingson, 2009) is useful to illustrate the approach we suggest. In addition, we argue for the necessity of allowing for multi-level exploration of data within mixed methods approaches. Particularly, we build on Wesler et al. (2008) who suggests computational social science researchers should aim to explore three levels of data. These are: 1) structural descriptions (i.e. patterns of interactions); 2) thin descriptions, which note the content of the interaction; and 3) thick descriptions, to provide context and convey the meaning of the events by the participants.

In the next sections, we reflect on six methodological and analytical approaches we have employed in our research into learning in MOOCs. These constitute key approaches that can be used to analyse learner interactions in digital education settings. When addressing these, we discuss the particular affordances of each, highlighting how they offer different insights into the interaction and learning process. As such, these approaches can be employed independently or together with other approaches to obtain a more holistic view of learning. The six approaches are: description, structure, dialogue, typology, experience, and experimentation. A secondary objective of the reflexive account offered here is to encourage and foster needed methodological sensitivity and awareness in interdisciplinary MOOC research. In a recent systematic review of methodological approaches adopted in MOOC research, Raffaghelli et al. (2015), for example, identified a "lack of attention […] to the methodological aspects involved in this field of research" (p. 502), and found "little concern about explicitly declaring the research paradigms embraced" (p. 497).

**Description**

Many studies begin with a focus on ways to describe the phenomenon, in this case a massive open online course or set of courses. This is an important step, as understanding the course(s) being studied enables a way to situate and understand the findings, and generalise the results to other courses and contexts. A number of research questions can be asked at this stage, including, for example, 'what are the demographic characteristics of students that participate in MOOC discussion forums?'; 'what are the pedagogical aims of the course?' or 'what proportion of people pass the course?'.

While relatively straightforward, these questions are important and often form a foundation for later analysis. An understanding of these issues is crucial for how data is interpreted, what questions are asked, and what models are developed.

Research methods particularly suited for addressing these kinds of questions are descriptive analyses of digital trace data (e.g. examining the frequency of posts in a forum over the course), using visualisations, conducting pre and post surveys to collect demographic, motivation and satisfaction data; and some form of observation (see, for example, Belanger and Thornton, 2013; Deboer, Breslow, Stump and Seaton, 2013; Gillani et al., 2014a).

Initially each method can be used independently to explore the nature of the course and what happens over time. However, combining data sources can also be useful, e.g. the use of the survey data to see the educational level of the people passing the course; or linking posting behaviour with observation data about what is happening at the time in the course (e.g. spikes in participation related to project milestones). In our own work, the patterns from the digital trace data echoed our qualitative understandings obtained through our observation of the course, and the research we had done to understand the course design and objectives (Gillani et al., 2014b). Thus, in the design of the research, being able to link data sets (e.g. a survey response with the respondent's digital trace data – e.g., their forum posts, video views, and other actions) can be very useful and worth incorporating into the research design where possible.

Two common challenges that need to be addressed are firstly, how to define certain variables. This is a challenge because these are settings where people do not need to learn at regular times or complete the course (or even begin it once that have signed up). Therefore, defining important aspects of the learning process requires careful consideration. For example, DeBoer and colleagues offer approaches as to how enrollment, participation, curriculum, and achievement can be measured in such settings – which go beyond more commonplace definitions that are used in more traditional learning environments (De Boer et al., 2014). A second significant issue is the poor response rate to pre and post course surveys, which are often used to collect information about demographics, motivation and satisfaction of learners. Frequently, these surveys have less than one in ten course participants (however defined) completing this information. While the numbers of respondents are large, the response rates are low, and are therefore likely to suffer from significant biases. Such data would typically be considered too weak in social science research to be valid, yet it is currently accepted in MOOC research – which is clearly problematic.

**Structural connections**

While the descriptive analyses of a course is valuable, it is important, when the focus of the research is on learner communication and interaction, to explore questions about who is talking to whom, and how information spreads through the forums. Indeed, there are a variety of techniques that can be used to analyse the more structural aspects of the forum. Here we review approaches that we have employed: the use of social network analysis – and ways to determine the significance and vulnerability of these social networks, and the use of models of social contagion to examine information flow between course participants.

Network analysis has exploded in recent years as a method of investigating how individual actors – including those in educational contexts – interact with one another (Easley and Kleinberg, 2010, Rabbany et al., 2014). Social network analysis (SNA) in particular helps model the spatially and temporally influenced social relationships (edges) between individuals (nodes). From such analysis it is possible to understand who is talking to whom in a MOOC and how these interactions develop and change over time. Previous studies in education have leveraged these techniques, albeit with small-scale datasets (Palonen and Hakkarainen, 2000; Cho, Gay, Davidson, and Ingraffea, 2007). The rise of "big data" and the tools that enable its analysis have encouraged more recent large-scale investigations that leverage the theory and practice of SNA in learning contexts (Kossinets and Watts, 2006; Vaquero and Cebrian, 2013).

When utilising SNA a number of key decisions need to be made in defining network structure – particularly because subsequent analysis of these networks is largely dependent on these modelling decisions. For example, in our own research, we defined nodes in the network to represent learners that created at least one post or comment in a discussion thread; an edge connected two learners simply if they co-posted in at least one discussion thread. Thus, we kept our definitions relatively simple: participation was defined as posting text to the discussion forum, and connections between learners were conceptualised as undirected (i.e., we assumed no directional flow of information, and instead allowed the connections between nodes to represent the *potential* for information to be shared between any two learners). A consequence of this approach was that we did not account for viewing (or 'lurking') behaviour, which is an important part of forums (Preece, Nonnecke, and Andrews, 2004). Equally, we did not compute who spoke to whom and therefore there was no obvious way for us to discern which way information was transmitted between actors despite this aspect being clearly important for learning. However, others have worked with directed connections within MOOC work (Yang et al., 2013).

Time is another variable that influences the creation of the network. Taking slices of time (e.g. a week) for the building of a network is one approach (Gillani and Eynon, 2014). This might be problematic, though, as it is an artificial time frame imposed by the research team rather than defined by those participating in the forums themselves (Krings et al., 2012). However, assigning no constraints on the network is problematic as well as using large time-intervals (e.g., the entirety of the course) renders the "thread network" visualizations very dense, and thus difficult to interpret. Given these issues, it is important to be aware of the choices that are being made in the creation of the network, and what ramifications these have on subsequent analysis.

While SNA provides valuable understanding of online social networks for learning (Haythornthwaite, 1996; 2002; 2009); networks only tell a partial story. This is not least because of the fact that not all links generated are equally important, and two learners' co-participation in a thread is not necessarily indicative of a meaningful social exchange. For example, a lot of exchanges might be simple introductions, or requests for help, or thoughtful reflections on the course material. These interactions have different implications for learning: some are irrelevant, others are meaningful.

The way we addressed this challenge was to conceptualise the observed communication network in each course sub-forum as a noise-corrupted version of the "true" network – i.e., one that depicts meaningful communication between students (Gillani et al., 2014b). Inspired by methods from machine learning (Psorakis et al., 2011), we then generated a set of "sample" communication networks based on the trends in the network we constructed, and tested for the likelihood that any given link in the observed network was present by chance, instead of indicative of a statistically significant interaction. Interestingly, some sub-forums retained more links than others, and these corresponded with those we have identified through our qualitative observation as venues facilitating meaningful interaction (Gillani et al., 2014a).

Another way to understand the structure of the forums is to examine the vulnerability of the networks. Vulnerability of networks has been studied across disciplines (Holme et al., 2002). For example, power systems engineers often ask which "critical set" of network components must be damaged in a functioning circuit in order to cut off the supply of electricity to the remaining nodes (Albert et al., 2000). Thus, it is possible to ask a similar question from an educational perspective: which "critical set" of learners is responsible for potential information flow in a communication network - and what would happen to online discussions if the learners comprising this set were removed? Vulnerability can be defined as the proportion of nodes that must be disconnected from the network in order to rapidly degrade the relative size of the largest connected component to the total number of nodes.

Intuitively, the vulnerability of MOOC discussion networks indicates how integrated and inclusive communication is. Discussion forums with fleeting participation tend to have a small proportion of very vocal participants comprise this set: removing these learners from the online discussions would rapidly eliminate the potential of discussion and information flow between the other participants. Conversely, forums that encourage repeated engagement and in-depth discussion among participants have a proportionally larger critical set, and discussion is distributed across a wide range of learners. By analysing vulnerability in different sub-forums, it is possible to determine how group communication dynamics differ according to the topics being discussed; and similar to the techniques described above, those sub-forums that were identified as less vulnerable were also identified as such in our qualitative data (for full details of the methodology see Gillani et al., 2014a).

A complementary approach to thinking about the structure of MOOCs forums is to explore how information spreads in these networks. Doing so may ultimately reveal how forum participation promotes knowledge construction. In our work we investigated this approach using an information diffusion model similar to the Susceptible-Infected (SI) model of contagion (Kermack and McKendrick, 1972) which has been extensively used in previous work to model social contagion (Onella et al., 2007). Although very simplistic, the SI model is very useful in analysing the topological and temporal effects on networked communication systems; and enabled comparison of information spread within different networks (see Gillani et al., 2014b for full details).

While these are just some of the techniques that can be used to explore the structure of MOOC forums, it is clear that the use of the digital trace data can provide a very important 'layer' of information about learning, yet is more powerful when used in combination with other methods. We now turn to looking at another aspect – dialogue.

**Examining dialogue**

When examining interaction and learning, it is not just the structures of the communication that are important: the content of what is being said is of great importance, too. Indeed, the role of dialogue and discourse in the learning process is recognised in a number of learning theories in different ways including in the work of Pask, Papert and Vygotsky (Ravenscroft, 2003). Thus techniques to address questions such as 'what is being discussed?', or 'what kinds of feelings are being conveyed?' or 'what knowledge construction is occurring in the forums?' are important for MOOC research.

Methodologically, there are a number of approaches to analysing online interactions. Content or discourse analysis has been used in previous higher education research in order to understand learner interactions that take place online (e.g. Stahl et al., 2006; De Weaver, Schellens, Valcke, and Van Keer, 2006; Gunawardena, Lowe, and Anderson, 1997). Often, these discourse analyses were conducted by the research teams, but the scale of MOOC discussion forums makes this difficult. Indeed, while it has been attempted in some previous work with MOOCs including our own (Gillani et al., 2014a) many researchers opt for more automated data coding approaches drawing on fields of text mining, natural language processing and computational linguistics. For example Wen and colleagues used sentiment analysis to determine affect while learning in MOOC forums to assist with understanding drop out (Wen et al., 2014). Others, beside using automated algorithms, have also crowdsourced their data analysis, for example using Mechanical Turk, to categorise Speech Acts in MOOC forums (Arguello and Shaffer, 2015).

Regardless of the approach, a number of decisions have to be made when coding such data. This ranges from the unit of dialogue and analysis (e.g. the word, the sentence or the entire response), the 'human' versus 'machine' elements of data coding (e.g. what proposition of codes need to be examined by people), to what precisely is being coded. In our work, we selected the response as the unit of analysis; all codes where 'human coded' and we employed a thorough coding scheme that aimed to measure a number of dimensions. The dimensions included a focus on the level of knowledge construction (e.g. ranging from no learning, through to four types of sharing and comparing of information, to more advanced stages of knowledge construction such as negotiation of meaning, (Gunawardena et al., 1997)); the communicative intent in the forums, selecting from five categories: argumentative, responsive, informative, elicitative and imperative (Clark et al., 2007; Erkens and Jenssen, 2008); and the topic of the post. It is important to note that coding schemes should not only be developed in correspondence with the theories of learning guiding the particular research project, but also, after preliminary observations of online course discussions to account for the nuances of any particular learning setting.

Once this data is collected and analysed, it can be used as an input into several different models for subsequent analysis. We used our coded forum data to create a typology of learners, which we describe next.

**Interpretative models**

From the descriptive, structural and dialogue approaches to analysing MOOC data we collect a great deal of valuable information. However, for the most part the questions these techniques can answer when used in isolation remain at a relatively descriptive level. Nonetheless, these data can be used in combination to provide more interpretative analyses of the learning and interaction that goes on in these contexts.

These can range from questions, such as, 'how does participation in discussion forums relate to students' final scores?'. This might, for example, be examined using statistical techniques (e.g. cross-tabulations or ANOVA) by relating posting in the forum to final scores, as compared to other activities, such as viewing the lecture videos; or by examining the relationships between demographics and outcomes. One way of achieving this is through relatively simple models of participation (i.e. did a person post or not, or did they post frequently or not (Davies and Graff, 2005)), or the model could take into account the network structure in some respect. For example, Vaquero and Cebrian took such an approach when examining whether high-performing students tend to interact with other high-performing students in online learning settings (Vaquero and Cebrian, 2013). Essentially, these analytical strategies aim to combine rich data sources in creative ways in order to build robust models that account for the complexities of learning.

In our own research we achieved this by first creating a typology of learners based on the content of their posts. While previous studies in education have opted for clustering approaches such as K-means or agglomerative methods (e.g., Ayers et al., 2009), we chose Bayesian Non-negative Matrix Factorization (BNMF) because it afforded a modelling flexibility and robustness that was better-suited for this particular dataset and application domain. This analysis allowed us to identify distinct groups of learners. We then connected this data to other data points that we collected from the descriptive and structural analysis; i.e. demographics (education, country and age), posts and views patterns in the forums, and outcomes (i.e. whether they submitted a final assignment and then passed or failed the course). This provided us with a useful typology of learners that fit with existing theoretical models of learning and education (see Gillani et al. (2014a) for full details).

**Understanding experience**

As highlighted in learning theories, experience is a really important aspect of understanding the learning process. Qualitative methods, such as observation and interviews, tend to be highly appropriate approaches to gather such data. There have been a number of qualitative studies of the MOOC experience; highlighting the practices that learners engage within outside the MOOC platform (Veletsianos, et al., 2015); and the experiences of 'lurkers' (Kop, 2011). This follows a long tradition in online distance learning, and education more generally, where interviews have been a key approach to understanding the complexity of the student experience (Hara and Kling, 1999). Interviews can be carried out in a range of online settings and there has been a significant amount of research that has examined how different online platforms shape the interview data in different ways (O'Connor et al., 2008; James and Busher, 2009).

In our research, we interviewed participants primarily by Skype. The interviews were semi-structured, and the interview guide covered themes relating to learners' socioeconomic background, and educational and employment trajectories in addition to questions explicitly focusing on their MOOC engagement. This category of questions addressed learners' motivations for taking the course; their learning styles and preferences; their perceptions and uses of the course forums, and the significance of the forums for their overall experience; and their interactions with other course participants. These topics were discussed in relation to their current life circumstances and other MOOCs they might have taken. Interviews were transcribed prior to the analysis and were conducted to provide data primarily about the motivations and experiences of learning that was not possible to obtain from the other methods in this project (Eynon et al., 2014).

One of the key challenges of carrying out interviews in large scale settings is trying to obtain some form of meaningful sample that is purposive / theoretical. While qualitative sampling does not rely on quantity, it is difficult to know precisely what kinds of experiences are being captured when interview studies may typically only focus on 30 participants amongst the tens of thousands of learners who originally signed up for a course. In our study for example, because we simply spoke to anyone who was interested in speaking to us, our interviewees for the large part were clearly some of the most committed learners; most MOOC participants end up disengaging from the course, and so, are also unlikely to participate in an interview; thus are unlikely to be "typical" MOOC learners. The use of carefully targeted invites, and appropriate incentives, are one way to deal with these issues. However, because researchers tend to know very little about the backgrounds of MOOC participants, a purposive sample is difficult. In our work, we were able to map our participants onto the quantitative typology (above) and this provided us with a better understanding of who we had spoken to. This method also enabled another very valuable way to combine 'small data' (data gleaned from participant interviews) with 'big data' (macro-level interaction patterns and other course-wide trends). Such an approach could be used in future work as a way to identify additional participants to interview. It may also help cross-validate findings and provide deeper insights than possible through leveraging 'small data' or 'big data' alone.

**Experimentation**

As understanding of MOOCs begins to develop, a number of researchers are beginning to focus on more experimental research methods to be able to make more causal claims and determine interventions that may positively support the learning process (Chudzicki et al., 2015; Reich, 2014). While these kinds of online field experiments are easier to conduct than similar experiments in the classroom setting, they are not without their methodological challenges (Lamb et al., 2015).

In our own work we conducted an email intervention campaign to explore how different discussion thread recommendation emails can promote social engagement among participants in MOOCs. Emails invited learners in an online business course to participate in group discussions by linking to a sampled set of active discussion threads in the course's forum. These emails were sent each week to between 30,000 and 45,000 course participants – totalling nearly 200,000 emails sent during the 5-week course. Course participants were randomly assigned to an email "treatment" group at the beginning of the course and remained in that group for the entirety of the course. Treatment groups for each weekly email campaign were determined by toggling 3 experimental variables: the type of email introduction (social vs normal; social emails mentioned the names of a few other forum participants); the type of email body (with content previews of five threads / without preview); and the method used to select the discussion threads included in each email (random; random excluding introductory posts; most popular, i.e. largest number of posts; and highest reputation, i.e. threads with participants that had largest number of upvotes). Overall, there were 16 possible email treatment groups.

Based on analysis of basic engagement statistics per recipient (namely, email opens and click-through rates on hyperlinked discussion threads), and posts and views in the forums, we found that emails containing the names of other forum participants had lower click-through rates than emails without any social information. Moreover, we found that discussion threads selected based on user reputation yielded higher click-through rates than those selected based on overall popularity. Email open rates remained high (approximately 30-40%) across various treatment groups throughout the 5-week intervention, suggesting interest from course participants in keeping up with what was being discussed in the forums.

This method also has its own limitations, namely the rather low click-through rates over all treatment groups (on average, less than 5%), as well as the lack of a consistent control group across all five weeks of the intervention – which diminished our ability to compare subsequent forum activity for those that received emails and those that were never sent any emails. Post-course survey responses, however, suggested that the discussion thread recommendation emails played an important role in reminding people about the discussion forums. This insight implies that further investigations – including participant surveys – will shed additional light on how email campaigns may help "nudge" participants towards more meaningful interactions and deeper engagement in massive-scale learning settings.

**Future research**

As is clear from the section above, there are multiple ways to combine data mining techniques with a wide array of social science methods to shed light on how people are interacting and learning in MOOCs; and we suggest that this is the most appropriate way to really understand what is happening when people learn in these large-scale semi-formal settings.

An important additional issue to raise concerns ethics, particularly given the fast pace of change in this area, where our uses of technology for learning and research changes faster than legal or institutional frameworks. In such contexts ethical committees alone should not be relied upon, and both researchers and practitioners also have responsibilities to consider in order to stand by their ethical decisions and codes of practice (Henderson et al., 2013, Pring, 2001).

Within MOOC research, there has been a great deal of debate in recent years around issues of privacy, with the rise of educational data mining and learning analytics (Pardo and Siemens, 2014). This is in part related to the different stakeholders involved, with computer scientists and social scientists working within quite different ethical codes, and the tensions in some cases between commercial and academic codes of conduct in research and practice (Marshall, 2014). However, this is not solely an issue for learning analytics and educational data mining

– ethical issues in online qualitative research in learning and education are also a continuing challenge and deserve attention (Kanuka and Anderson, 2007).

While it is not the purpose of this chapter to debate these ethical issues in depth, we would encourage all current and future researchers engaged in this area of work to consider a range of debates, particularly when combining a different range of data sources together. Valuable texts include Eynon et al., (2008, 2016); Slade and Prinsloo (2013); Markham and Buchanan (2012). Ultimately, we would recommend that researchers navigate the terrain according to their own epistemological frameworks, with an awareness of the current debates and a commitment to contributing to it.

As the hype around MOOCs begins to fall away, research opportunities in this area remain very rich both for online education and beyond. The findings from studies on crowd-based learning are likely to be applicable and transferrable to a whole range of settings where online crowds come together to achieve certain goals, from citizen science to political participation. Using theoretical perspectives from learning and education provide a valuable lens to many of these contexts. However, whether the focus is on MOOCs, crowdsourcing or the next online learning innovation, researchers must continue to build on what has gone before. Learning is messy and difficult to measure, occurring both within and across individuals in a range of contexts across their course of life. We suggest that data mining or qualitative investigations alone will never be sufficient to understand this complex process, and that significant value lies in combining these methods for a more robust, holistic understanding of how people learn.